\documentclass[prl,twocolumn,superscriptaddress]{revtex4}

\usepackage[utf8]{inputenc}
\usepackage[english]{babel}
\usepackage{amsmath,amssymb,graphicx,bm}

\usepackage{graphicx} 
\usepackage{subfigure} 
\usepackage{color}

\newcommand{\beqal}[1]{\begin{eqnarray}
\label{#1}}
\newcommand{\eeqa}{\end{eqnarray}}

\begin{document}

\newcommand{\vesp}[1]{\left\langle#1\right\rangle} 
\title{Paths towards equilibrium in molecular systems: a mesoscopic Mpemba-like effect in water}

\author{A.~Gij\'{o}n}
\affiliation{Instituto de Ciencia de Materiales 
de Madrid (ICMM--CSIC),
Campus de Cantoblanco, 28049 Madrid,
Spain}
\author{A.~Lasanta} 
\thanks{email: alasanta@ing.uc3m.es}
\affiliation{G. Mill\'{a}n Institute, Fluid Dynamics, Nanoscience and Industrial Mathematics
Department of Materials Science and Engineering and Chemical Engineering
Universidad Carlos III de Madrid 
Legan\'{e}s, Spain}

\author{E.~R.~Hern\'{a}ndez}
\thanks{email: Eduardo.Hernandez@csic.es} 
\affiliation{Instituto de Ciencia de Materiales 
de Madrid (ICMM--CSIC),
Campus de Cantoblanco, 28049 Madrid,
Spain}

\keywords{Non-equilibrium thermodynamics $|$ Mpemba effect $|$ Memory effects in condensed matter } 

\begin{abstract}


The so-called Mpemba effect, i.e. the observation that the warmer of two otherwise identical systems cools faster when both
are refrigerated in the same thermal reservoir, is a hotly debated topic in condensed mater physics and statistical mechanics. Although it 
has been found in several non-equilibrium model systems, its very existence in water, the system in which it has been historically reported, 
is still open to question. Here we show using numerical simulations that a Mpemba effect is indeed
present in water. 
We find that the effect occurs when equipartition of energy is not present in the initial state. Interestingly, the
effect is observed without the intervention of a phase transition, and it is therefore seen to be a purely non-equilibrium relaxation effect.


\end{abstract}

\date{\today}

\maketitle


An unperturbed out-of-equilibrium system will evolve in time until it finally reaches a state of equilibrium, from which it 
will not spontaneously depart. This evolution is system-specific, and dependent on the initial conditions. Experimental
observations~\cite{Keim} reveal that frequently the evolution towards equilibrium depends on the system's thermal history, a phenomenon
known as a memory effect. Memory effects are common in condensed matter systems~\cite{Kovacs,Prados,Rodriguez,Xie,Fiocco,Schoenholz,Nagel}, having been documented in disordered 
materials~\cite{Lahini}, spin glasses~\cite{Jonason}, granular matter~\cite{Prados},  polymers~\cite{Struik}, biological
systems~\cite{Kursten} and batteries~\cite{Sasaki}; 
they are thus of fundamental as well as practical interest, but our understanding of them is still incomplete. 
%
%
An intriguing memory effect named after Mpemba~\cite{Mpemba} 
is frequently ascribed to water~\cite{Ball,Ouellette}; it refers to the observation that the warmer of two otherwise identical beakers of water, 
when put in contact with the same thermal reservoir, may cool faster under 
certain conditions. Even though it has been described since antiquity~\cite{Aristotle} the Mpemba effect~(ME) remains controversial to date. Experimental 
efforts have thus far been unable to provide a conclusive and reproducible picture of the effect~\cite{Brownridge,Burridge}.
The phenomenon is not specific to water; indeed the earliest modern reference to 
it~\cite{Mpemba} describes how it was first observed in ice-cream mixtures; there is also report of its observation in clathrate 
hydrates~\cite{Ahn}.  A number of computational experiments have reported Mpemba-like
phenomena in various idealised model systems, such as in granular fluids~\cite{Lasanta17,Torrente19}, spin glasses~\cite{Baiti18} and Markovian systems~\cite{Lu17,Klich18}. 

Here we describe a hitherto unacknowledged non-equilibrium Mpemba-like effect in atomistic models of bulk water, which we
refer to as the {\em mesoscopic Mpemba\/} effect. The term {\em mesoscopic\/} is used here in the sense of 
Van~Kampen~\cite{vanKampen}, i.e. a description of the thermal evolution of a system in terms of macroscopic 
variables (e.g. temperature) that are nevertheless affected by fluctuations resulting from the microscopic dynamics of its
constituents. We also use the term to distinguish our observations from the conventional, macroscopic ME, which, if real, 
takes place over time scales much longer than can be observed in the effect we report here.

\begin{figure*}[th]
\centering
\includegraphics[width=15cm, keepaspectratio]{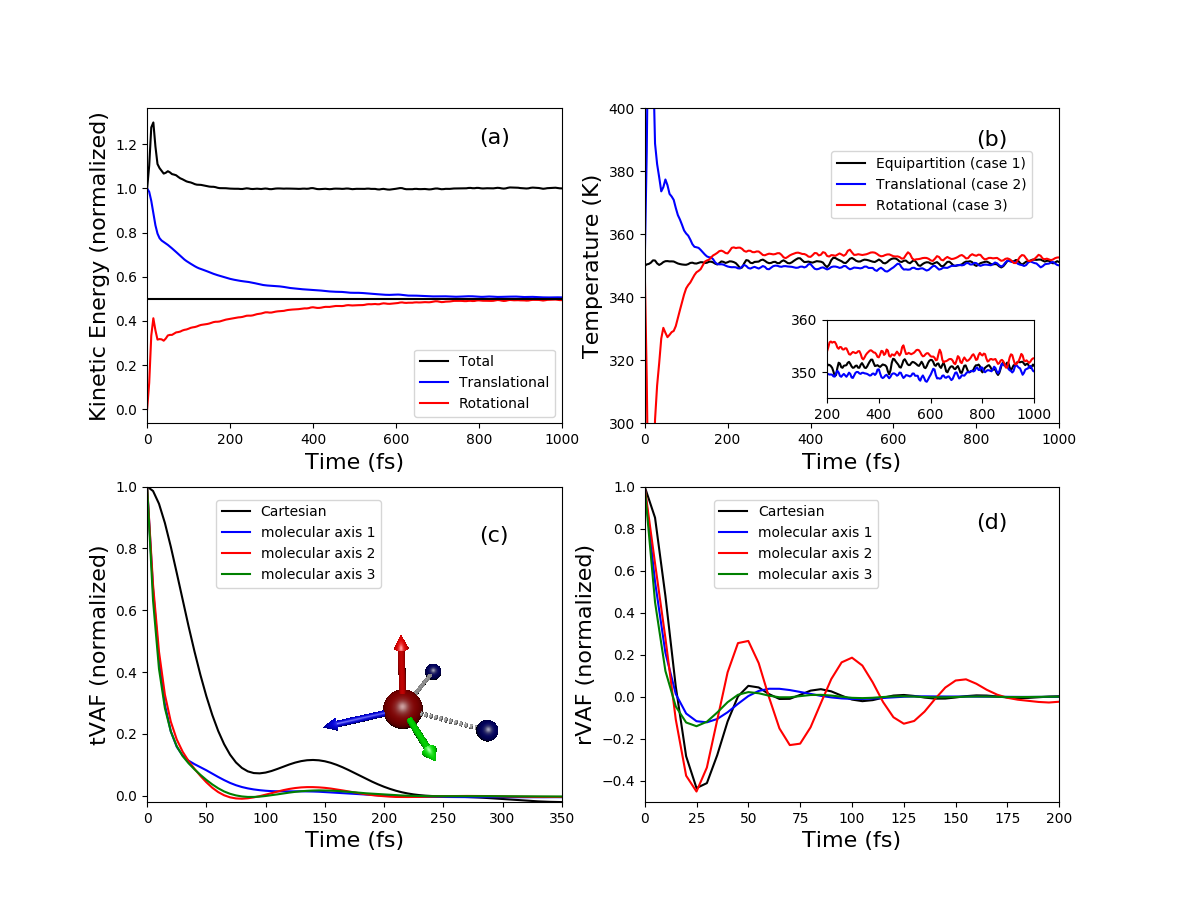}
\caption{(Color online) (a) Evolution of the total, translational and rotational kinetic energy towards equipartition for liquid TIP4P water; 
at the start of this simulation
all the kinetic energy was in translational modes; (b) evolution of the temperature in three different situations: (1) starting from equipartition, (2) kinetic energy initially only in translational modes (the case displayed in panel (a)), and (3) kinetic energy initially in rotational modes only (the 
reverse case of that shown in panel (a)).  (c) Translational velocity auto-correlation function (tVAF) of liquid TIP4P water at 350~K. VAFs of Cartesian velocity components are shown in black, while those of velocity components in the molecular frame are shown in blue, red and green; the disposition of
the molecular frame is indicated by the inset figure, in which each axis is coloured to accord with the corresponding tVAF component. (d) 
Rotational velocity auto-correlation function (rVAF) of liquid TIP4P water at 350~K. The colour coding is the same as for panel (c).  As can be seen, there are appreciable differences between the different molecular-frame components
(both translational and rotational) as well as between the translational and rotational mode dynamics.}
\label{fig:NVE}
\end{figure*}

In our simulations we monitor the process of equilibration from carefully-prepared initial conditions in water systems. To this end, 
we have conducted constant-volume molecular dynamics~(MD) simulations of systems containing~64000 water molecules
in periodic boundary conditions at densities corresponding to liquid water and hexagonal ice. Simulations were carried out
using two different implementations of molecular dynamics, NAMD~\cite{NAMD} and LAMMPS~\cite{LAMMPS}, both giving the same results.  
In order to insure that the effects observed were not an artefact of a 
particular water model, we have used several different standard models, including the SPC, TIP3P
and TIP4P~\cite{TIP4P} rigid models, as well as a flexible version of the TIP3P model~\cite{TIP3P}.
The equations of motion were integrated using a time-step of 1~fs.
Initial atomic velocities were generated by sampling the Maxwell-Boltzmann
distribution at the desired target temperature. Starting configurations for the liquid with rigid models were obtained 
from a simple-cubic lattice distribution of molecules with a density of 1~$\mbox{g/cm}^3$. These configurations were then
subjected to an equilibration run during which velocities were re-scaled to drive the system towards the desired temperature, lasting
a total of 10~ps. Production runs were 5~ps long. Runs with the flexible TIP3P model used as initial seed a previously equilibrated 
configuration of the rigid model, which was subsequently equilibrated with velocity re-scaling for 200~ps, followed by a production 
run of 50~ps. For the solid phase, we started from a supercell of ice Ih containing~64000 molecules, constructed so as to have no
net dipole; this was then subjected to the same equilibration procedure as the liquid phase. Once a given system had been equilibrated
at the desired temperature, we took a representative configuration and constructed new, out-of-equilibrium initial conditions by 
modifying the velocity distribution of the molecules, so as to study the system's evolution back to equilibrium. We considered the 
following four cases: case~1, in which the velocities were unaffected (normal equilibrium case); case~2, in which the kinetic energy was
placed only in translational modes (molecular rotations were initially frozen); case~3, in which the translational modes were initially frozen, with the kinetic energy placed in rotational modes, and finally case~4 in which both translation and rotation modes were initially frozen, and the kinetic energy was contained in internal molecular vibrations (this is only relevant for the flexible TIP3P model). Clearly, cases~2 to~4 above 
are very extreme and particular and are only used as convenient starting conditions from which to monitor the process of equilibration; they are far from equilibrium velocity distributions; our conclusions below do not depend on such starting conditions. {\em The use of thermostats to impose a target temperature was explicitly avoided to prevent any 
interference with the intrinsic relaxation dynamics of the system\/}. 


In Fig.~(\ref{fig:NVE}) we display a sample of the observations obtained from the simulations described above.
The system is liquid water modelled with the rigid TIP4P~\cite{TIP4P} model, starting from a configuration
resulting from a long equilibration period at 350~K. This model, being rigid, has only translational and rotational degrees of freedom. After the equilibration period, the velocities of atoms in the system are modified such that all
the kinetic energy is put in molecular translational modes, while that in rotational modes is set to zero (case~2 above). 
This distribution obviously breaks the 
equipartition of kinetic energy among different modes that one expects to find under conditions of thermal equilibrium, and
Fig.~(\ref{fig:NVE}a) shows how the system evolves towards restoring equipartition between 
translation and rotation in this case. Fig.~(\ref{fig:NVE}b) displays the evolution 
of the instantaneous temperature of the system in three different cases, namely, the case considered in Fig.~(\ref{fig:NVE}a) (case~2), 
the alternative case in which the initial kinetic energy is put only in molecular rotational modes (case~3 above), 
and normal equilibrium (case~1). As can be appreciated in Fig.~(\ref{fig:NVE}a), the translational
and rotational kinetic energies evolve towards their equipartition value at slightly different rates. Indeed, the translational energy is seen to reach
its equipartition value faster than the  rotational one. This is also the case when the initial conditions are inverted so that all kinetic energy is
initially in rotational modes, i.e. the translational kinetic energy reaches its equipartition value before the rotational one, only in that case it 
reaches it from below. The different rates of evolution towards equipartition of the two kinds of modes 
present in this system result in a transient reduction (or increase, in case~3) of the total kinetic energy 
with respect to its equilibrium average value. Although this difference with 
respect to the equilibrium value is small, it nevertheless translates into a noticeable (albeit transient) temperature mismatch
with respect to the equilibrium value of approximately $\pm5$~K, as can be appreciated in Fig.~(\ref{fig:NVE}b).

The behaviour observed in Figs.~(\ref{fig:NVE}a,b) is not so surprising if one considers that different kinds of degrees of freedom experience different dynamics, as made evident by the corresponding velocity auto-correlation functions (VAF), displayed in Fig.~(\ref{fig:NVE}c,d).  
Fig.~(\ref{fig:NVE}c) plots the VAFs of the centre-of-mass translational velocity (tVAF), while Fig.~(\ref{fig:NVE}d) displays the VAFs of the rotational velocity (rVAF). 
It can be observed that the translational and
rotational VAFs are very different from each other. 

\begin{figure*}[ht]
\centering
\includegraphics[width=15cm, keepaspectratio]{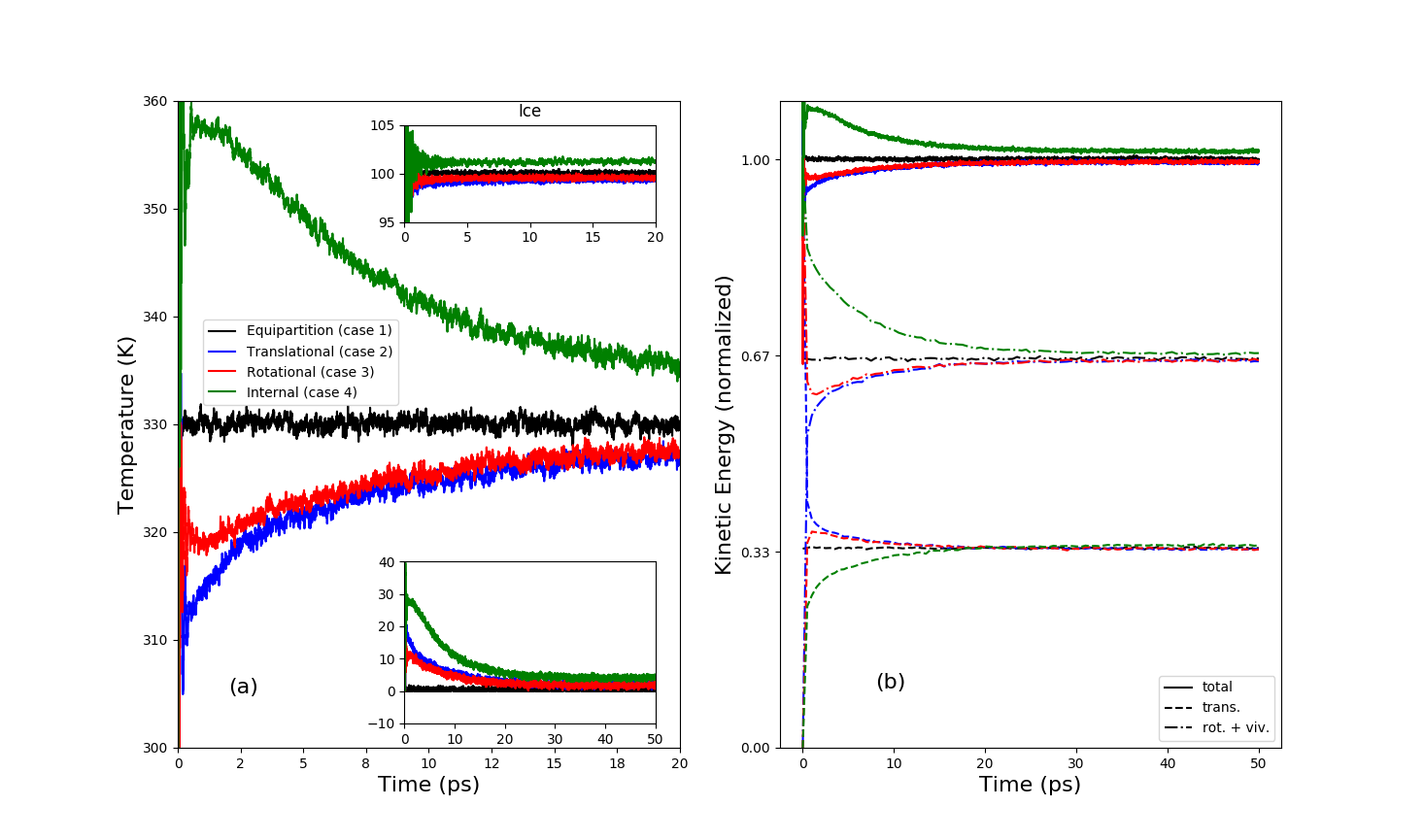}
\caption{(Color online) (a) Instantaneous temperature evolution for flexible TIP3P water; the black curve shows the temperature for a system
in equilibrium at 330~K (case~1); case~2 (blue line) is the temperature evolution for a system of the same size in which all the kinetic
energy is initially put in translational modes, while in case~3 (red) the kinetic energy is initially concentrated in rotational modes only, 
and in case~4 (green) the kinetic energy is initially distributed in internal (vibrational) modes only. The lower inset shows the evolution of 
$\left | T-T_e \right |$, being $T_e$ the equilibrium temperature, where the different relaxation times can be clearly observed  in the range up to 50~ps; the upper inset shows the temperature evolution obtained in similar simulations for the 
case of ice with a total kinetic energy corresponding to an equilibrium temperature of 100~K (see Supplementary Information file).
Panel (b) shows the decomposition of the kinetic energy into translational and rotation plus vibrational contributions for the four cases 
displayed in panel (a), with the same colour coding. }
\label{fig:kinetic}
\end{figure*}

We next consider the case in which a water molecular model with internal degrees of freedom is employed. Specifically, we consider the 
flexible TIP3P~\cite{TIP3P} model, which incorporates harmonic potentials for both the \mbox{O-H} bonds and the \mbox{H-O-H} bond-angle. Strictly speaking, 
quantum effects in the molecular internal degrees of freedom are more prominent than in translation and rotational motion, and thus 
equipartition does not hold for such modes in the real situation, but in line with standard 
practice in molecular mechanics simulations we are going to obviate this fact. Due to the
different time-scales involved in the dynamics of internal degrees of freedom and molecular translation/rotation modes it can be anticipated that 
energy exchange between the first and the latter will be comparatively slow. We can thus expect that the timescale to reach equilibrium (equipartition
of energy between the different modes) will be longer than for rigid water models. Fig.~(\ref{fig:kinetic}) shows that this is indeed the case;
displayed in Fig.~(\ref{fig:kinetic}a) is the temperature evolution for flexible TIP3P water at 330~K in four different cases: 
equilibrium, all kinetic energy
initially placed in purely molecular translational modes, all kinetic energy initially placed in rotational modes, and all kinetic
energy initially distributed among internal molecular vibrations; the latter case is only considered for completeness since, as noted above, quantum
effects will be more important in the internal degrees of freedom. Therefore, we focus our discussion on the second and third cases. The upper inset in panel~(\ref{fig:kinetic}a) shows similar results for the case
of  TIP3P ice at 100~K, demonstrating that similar effects can be observed in out-of-equilibrium solids as well as in liquids (more
details of the latter simulations can be found in the Supplementary Information file).
Panel~(\ref{fig:kinetic}b) shows the distribution of kinetic energy into different modes for all the simulations of the liquid phase considered in Fig.(\ref{fig:kinetic}a);
as can be seen there, equipartition still holds for the equilibrium case, with 
translational kinetic energy accounting for 1/3 of the total kinetic energy, and rotational plus vibrational kinetic energy for the remaining 2/3. 
In the cases of initial non-equilibrium distributions of kinetic energy, equipartition is attained in due course, but much more slowly than 
when a rigid molecular model is used, see Fig.~(\ref{fig:NVE}a). As can be appreciated in Fig.~(\ref{fig:kinetic}a), when the kinetic energy is 
initially placed exclusively in either translational or rotational modes
the instantaneous temperature approaches the equilibrium value of 330~K from below,  while if only internal molecular vibrations are initially
excited, the approach to the equilibrium temperature is from above. In all three cases, however,   equilibration is attained over a time scale 
that is at least an order of magnitude longer than in the case of rigid water molecules.

In conclusion, the results displayed in Figs.~(\ref{fig:kinetic}) reveal a mechanism that we refer to as
the mesoscopic ME.  If equipartition does not hold, and in particular if there is an excess of kinetic energy 
contained in either (or both) translational and rotational modes, the resulting temperature would be lower, by up to a few degrees, to that of 
an identical sample obeying equipartition. This is a reflection of the different time-scales associated to the dynamics of different degrees of freedom. Equipartition is only ensured when the sample is in conditions of thermal equilibrium, but breaking
that equilibrium (for example, by placing the sample in a colder environment, as is done in Mpemba effect experiments) could result in the 
breaking of equipartition. Interestingly, the results reported here show the possibility of the existence of an inverse ME in water, as far as we know not observed or predicted, and which is present in other non-equilibrium systems ~\cite{Lasanta17,Torrente19,Baiti18,Lu17}. It should be stressed that our results strongly support the hypothesis that the ME is an  effect that could arise in out of equilibrium relaxation of complex systems.  The development of the multidimensional spectroscopy techniques,  allowing for the excitation of particular modes of specific water molecules and the visualisation of the response in the femptosecond timescale range~\cite{MUK00,COW07,HOCH07,ASH06} could open the door for the experimental validation of our hypothesis and results.

\section*{Acknowledgements}
{The authors thank Drs. R.~Ramirez, C.~P.~Herrero, V.~Mart\'{\i}n Mayor, F.~Barbero, Juan Jos\'e Lietor Santos and E.~J.~S.~Villase\~{n}or enlightening discussions.
The Galician Supercomputing Centre (CESGA) is thanked for access to their computational facilities. This work is funded by the Spanish {\em Ministerio de Econom\'{\i}a y Competitividad\/} through projects
FIS2-15-64222-C2-1-P and MTM2017-84446-C2-2-R.} 



\begin{thebibliography}{10}

\bibitem{Keim} N.~C.~Keim,  J.~Paulsen,  Z.~Zeravcic, S.~Sastry and S.~R.~Nagel, S.~R., arXiv:1810.08587v1 (2018).

\bibitem{Kovacs} A.~J.~Kovacs, J.~J.~Aklonis,  J.~M.~Hutchinson and A.~R.~Ramos,  J. Polym. Sci. Pt. B-Polym. Phys. {\bf 17}, 1097 (1979).

\bibitem{Prados} A.~Prados and E.~Trizac, Phys. Rev. Lett. {\bf 112} 198001 (2014).

\bibitem{Rodriguez} G.~F.~Rodriguez, G.~G.~Kenning and R.~Orbach, Phys. Rev. Lett. {\bf 91} 037203 (2003).

\bibitem{Xie} T.~Xie,  Nature {\bf 464}, 267 (2010).

\bibitem{Fiocco} D.~Fiocco, G.~Fo and S.~Sastry,  Phys. Rev. Lett. {\bf 112} 025702 (2014).

\bibitem{Schoenholz} S.~S.~Schoenholz, E.~D.~Cubuk, E.~Kaxiras and A.~J.~Liu, Proc. Natl. Acad. Sci. U.S.A. {\bf 114} 263 (2017).

\bibitem{Nagel} S.~R.~Nagel, Rev. Mod. Phys. {\bf 89}, 025002 (2017).

\bibitem{Lahini} Y.~Lahini, O.~A.~A.~Gottesman  and S.~M.~Rubinstein, Phys. Rev. Lett. {\bf 118} 085501 (2017).

\bibitem{Jonason} K.~V.~E.~Jonason,  J.~Hammann,  J.~.P.~Bouchaud and P.~Nordblad, Phys. Rev. Lett. {\bf 81} 3243 (1998).

\bibitem{Struik} C.~L.~E.~Struik, {\em Physical Ageing in Amorphous Polymers and Other Materials\/}, (Elsevier, Amsterdam, 1980).

\bibitem{Kursten} R.~Kursten,  V.~Sushkov and T.~Ihle, Phys. Rev. Lett. {\bf 119} 188001 (2017).

\bibitem{Sasaki} T.~Sasaki, Y.~Ukyo  and P.~Novak, Nature Materials {\bf 12} 569 (2013).


\bibitem{Mpemba} E.~B.~Mpemba and D.~G.~Osborne, Phys. Educ. {\bf 4}, 172 (1969).

\bibitem{Ball} P.~Ball,  Physics World {\bf 19} 19 (2006).

\bibitem{Ouellette} J.~Ouellette, Physics World {\bf 30} 20 (2017).

\bibitem{Aristotle} Aristotle. {\em Metaphysics\/}, (Clarendon, Oxford, UK, 1981).

\bibitem{Brownridge} J.~.D.~Brownridge, Am. J. Phys. {\bf 79} 78 (2011).

\bibitem{Burridge} H.~C.~Burridge and P.~F.~Linden, {\em Scientific Reports\/} {\bf 6} 37665 (2016).

\bibitem{Ahn} Y.-H.~Ahn, H.~Kang, D.-Y.~Koh and H.~Lee,  Korean J. Chem. Eng. {\bf 33} 1903 (2016).

\bibitem{Lasanta17} A.~Lasanta, F.~V.~Reyes, A.~Prados and A.~Santos, Phys. Rev. Lett. {\bf 119}, 148001 (2017).

\bibitem{Torrente19} A.~Torrente, M.~A. Lopez-Casta\~no, A.~Lasanta, F.~V.~Reyes, A.~Prados and A.~Santos, Phys. Rev. E (R) {\bf 99}, 060901 (2019).

\bibitem{Baiti18}  Janus col.: M.~Baity-Jesi, {\em et al\/}.  Proc. Natl. Acad. Sci. U.S.A (In press, 2019).

\bibitem{Lu17} Z.~Lu and  O.~Raz, Proc. Natl. Acad. Sci. U.S.A. {\bf 114}, 5083 (2017).

\bibitem{Klich18} I.~Klich and M.~Vucelja, arXiv:1812.11962.

\bibitem{vanKampen} N.~G.~Van Kampen, {\em Stochastic Processes in Physics and Chemistry\/}, (Elsevier, Amsterdam 2003).

\bibitem{NAMD} J.~C.~Phillips, R.~Braun, W.~Wang, J.~Gumbart,  E.~Tajkhorshid, E.~Villa, C.~Chipot,  R.~D.~Skeel, L.~Kale and K.~Schulten, J. Comput. Chem. {\bf 26} 1781, (2005).

\bibitem{LAMMPS} S.~Plimpton, J. Comp. Phys. {\bf 117}, 1 (1995).

\bibitem{TIP4P} W.~L.~Jorgensen, J.~Chandrasekhar, J.~D.~Madura, R.~W.~Impey and M.~L.~Klein,J. Chem. Phys. {\bf 79}, 926 (1983).

\bibitem{TIP3P} D.~J.~Price and ~C.~L.~Brooks, J. Chem. Phys. {\bf 121} 10096 (2004).

\bibitem{HOCH07} R. M. Hochstrasser, Proc. Nat. Acad. Sci. U. S. A. {\bf 104} 104190 (2007).

\bibitem{COW07} M. L Cowan et al., Nature {\bf 434} 199 (2005).

\bibitem{MUK00} S. Mukamel, Annu. Rev. Phys. Chem. {\bf 51} 691 (2000).

\bibitem{ASH06} S. Ashihara et al., Chem.  Phys. Lett. {\bf 424} 66 (2000).



%

\end{thebibliography}
\end{document}